\documentstyle[preprint,aps]{revtex}
\begin{document}
\draft




\title{Properties of Strangelets at Finite Temperature 
	\\in Liquid Drop Model
	\footnote{\it Partly supported by the National Natural 
Science Foundation of China}}
\author{Y. B. He$^{1,2}$, C. S. Gao$^{1,2,3}$, X. Q. Li$^{1,3,4}$,
 and W. Q. Chao$^{1,3,5}$}
\address{ 1. China Center of Advanced Science and Technology (World Laboratory),
        \\P.O. Box 8730, Beijing 100080,  China\\
	2. Department of Physics, Peking University,
         Beijing 100871, China\\
	3. Institute of Theoretical Physics, Academia Sinica, P.O. Box 
	2735, Beijing 100080, China\\
	4. Department of Physics, Nankai University, Tianjin 300071, China\\
	5. Institute of High Energy Physics, Academia Sinica, P.O. Box
	918(4),	Beijing 100039, China}
\date{\today}
\maketitle

\begin{abstract}
A comprehensive study of the properties of strangelets at zero and finite
temperature is presented within the framework of liquid drop model, including
the essential finite size effects. Strong parameter dependences of the 
properties are found and discussed. 
\end{abstract}

\narrowtext
\vfill

\newpage
\section{Introduction}

It was argued by 
Witten \cite{wit84} that strange quark matter might be the 
ground state of normal nuclear matter at zero temperature and zero pressure,
which was later supported by the studies of Farhi and Jaffe \cite{far84}
based on MIT bag model. 
The existence of stable strange quark matter would have some remarkable
consequences in cosmology and astrophysics. For instance, it has been suggested 
\cite{wit84} that some of the dark matter in the Universe could possibly 
exist in the form of strange quark matter which was produced when the 
Universe underwent the quark to hadron transition. The possible transition
of a neutron star into a strange star was discussed as well \cite{mad88,alc88}.
In addition, there have been proposals that
the strange stars might be the sources
of gamma rays with very high intensities \cite{alc86,hae91}.
Recently Greiner and his collaborators \cite{gre87}
have proposed that small lumps of strange quark matter (``strangelets'') might
be produced in ultrarelativistic heavy-ion collisions, and they could serve
as an unambiguous signature for the formation of quark-gluon plasma. In fact, 
several heavy-ion collision experiments at Brookhaven and CERN are searching 
for strangelets \cite{bar90}.

Most strange quark matter investigations \cite{far84,gre88,gil93}
have been performed based on shell model or liquid drop model.
While calculations in shell model are rather
tedious, recently Madsen \cite{mad94} has pointed out that the general 
structure of shell model results can be obtained more readily from liquid
drop model.
On the other hand, 
strange quark matter at finite temperature has attracted some special 
interests \cite{rei88,cha93}, since in some cases, such as ultrarelativistic
heavy-ion collisions, strange quark matter is expected to exist in a hot 
environment. However, these studies of strange quark matter at finite 
temperature have not completed the treatment of 
finite size effects, which are expected to play a substantial role in 
strangelets possibly appearing in the scope of ultrarelativistic heavy-ion
collision experiments \cite{mad93a}. 
The present work is an attempt to apply liquid drop model to strangelets 
at finite temperature, and could be a step towards the studies of
phase structure and evolution of strange quark matter.
We will include the important
finite size effects (volume, surface and curvature contributions) to study
the overall properties of strangelets at finite temperature, which might
give some clues to the search of strangelets.

In the literature \cite{far84,mad94,mad93} there have been discussions
on parameter dependences of the stability of strange quark matter at zero
temperature. However, it is interesting as well to study how the parameter
``windows'' for stable strange quark matter would become, if strange quark 
matter
at finite temperature, {\it e.g.} 
strange stars with a temperature up to tens of 
MeV, is assumed to exist stably in nature. In this work we shall investigate
the properties of strange quark matter at finite temperature for a wide range
of parameters.

\section{Formalism}
We consider the strangelet as a gas of up, down, strange quarks, their 
antiquarks, and gluons confined in an MIT bag. In liquid drop model
\cite{mad94}, the grand potential of the system is given by
\begin{eqnarray}
\Omega = \sum_i \Omega_i + BV.
\end{eqnarray}
Here $B$ is the bag constant, $V$ is the bag volume, 
and the grand potential of 
particle species $i$ can be written as
\begin{eqnarray}
\Omega_i=\mp T \int_0^{\infty} dk {dN_i \over dk} \ln[1 \pm 
\exp(-(\epsilon_i (k)-\mu_i)/T)],
\end{eqnarray}
where the upper sign is for fermions, the lower for bosons, $\mu$ and $T$
are chemical potential and temperature, $k$ and $\epsilon_i$ are 
particle momentum and energy. In a multiple reflection expansion \cite{bal70}
the density of states is given by
\begin{eqnarray}
{dN_i \over dk}=g_i\left\{ {1 \over 2\pi^2} k^2 V + f_S^{(i)}
\left({m_i \over k}\right)kS
+f_C^{(i)}\left({m_i \over k}\right)C+...\right\}.
\end{eqnarray}
For spherical strangelets $V={4 \over 3}\pi R^3$ is the volume of the bag,
$S=4\pi R^2$ is the surface area, and $C=8\pi R$ is the extrinsic curvature
of the bag surface. The factor $g_i$ is the statistical weight (6 for quarks
and antiquarks, and 16 for gluons). 

The surface term for quarks was given by 
Berger and Jaffe \cite{ber87} as
\begin{eqnarray}
f_S^{(q)}\left({m_q \over k}\right)=-{1 \over 8\pi}\left\{1-{2 \over \pi}
\arctan {k \over m_q}\right\},
\end{eqnarray}
In particular, the surface term for massless quarks and gluons is zero.
It has been shown by Madsen \cite{mad94} that the following ansatz works 
for the curvature term of massive quarks:
\begin{eqnarray}
f_C^{(q)}\left({m_q \over k}\right)={1 \over 12\pi^2}\left\{1-{3k \over 2m_q}
\left({\pi \over 2}-\arctan{k \over m_q}\right)\right\}.
\end{eqnarray}
For gluons the curvature term is \cite{bal70a}
\begin{eqnarray}
f_C^{(g)}=-{1 \over 6 \pi^2}.
\end{eqnarray}

After the construction of the grand potential as above, 
we can readily obtain
the thermodynamical quantities of the system as follows.
The net number of quarks, {\it i.e.} the number of quarks minus the number of 
antiquarks, can be derived from
\begin{eqnarray}
N_i=-\left({\partial \Omega_i \over \partial \mu_i}\right)_{T,V},
\end{eqnarray}
and 
\begin{eqnarray}
\label{con}
A={1 \over 3} \sum_i N_i
\end{eqnarray}
gives the total baryon number of the strangelet.
The free energy $F$ and internal energy $E$ of the strangelet are given by
\begin{eqnarray}
F=\sum_i (\Omega_i+N_i \mu_i)+BV,
\end{eqnarray}
and 
\begin{eqnarray}
\label{energy}
E=F+TS
\end{eqnarray}
with the entropy $S=-\left({\partial \Omega \over \partial T}\right)_{V,\mu}$.

\section{Numerical Results}
We can study the ground state properties of strangelets at finite 
temperature $T$, {\it i.e.} the lowest mass state for a given baryon number $A$
and temperature $T$, 
by minimizing the free energy $F$
with respect to the net number of quarks $N_q$, $N_s$ and 
the volume of the system $V$,
at fixed baryon number $A$.
However, for the sake of illustration, in Fig.~1 
we vary the strangeness fraction 
$f_s=N_s/A$ (net number of strange  quarks per baryon) instead of $N_s$, to 
show the minimum of the free energy per baryon. While the free energy
per baryon decreases as temperature rises up from zero to 30 MeV, the value of 
$f_s$ which minimizes the free energy per baryon changes little. Actually, as 
it will be shown later, 
the strangeness fraction of strangelet in its ground state
has weak dependence on temperature.

An easier way to investigate the ground state properties of strangelet is to 
minimize the free energy analytically with respect to $N_q$, $N_s$ and $V$,
under the constraint Eq.~(\ref{con}) for 
fixed $A$, which leads to the mechanical 
equilibrium condition
\begin{eqnarray}
\label{eq1}
-\sum_i {\partial \Omega_i \over \partial V}=B,
\end{eqnarray}
and the optimal composition condition
\begin{eqnarray}
\label{eq2}
\mu_q=\mu_s
\end{eqnarray}
(For details see Appendix).
Eq.~(\ref{eq1}) has the explanation that the pressure exerted by the quarks and
gluons in the bag is balanced by the bag pressure $B$. Eq.~(\ref{eq2}) might be
unexpected, however. It could be interpreted as follows. Since the total number
of quarks ($=N_q+N_s=3A$) in the strangelet 
is fixed, the free energy of the system is minimized 
when adding a massless quark (up or down quark) to the strangelet 
is energetically as favorable as 
adding a massive strange quark, 
{\it i.e.} $\mu_q=\mu_s$. Eq.~(\ref{eq2}) is usually
considered as a consequence of bulk strange matter at energy minimum. We argue
that Eq.~(\ref{eq2}) holds for strangelets as long as the liquid drop model
is applicable (applicability of liquid drop model to strangelets has been 
approved by Madsen \cite{mad94}). Nevertheless, if some other contributions,
{\it e.g.} Coulomb effects, are taken into account, the simple relationship in 
Eq.~(\ref{eq2}) will be modified. Eqs.~(\ref{con})$-$(\ref{eq2}) 
altogether specify
the ground state properties of strangelets.

A strangelet at zero temperature is stable relative to $^{56}$Fe nuclei if its
energy par baryon, $E/A$, is less than 930 MeV. At finite temperature, however,
a strangelet can radiate elementary particles like nucleons and pions 
from its surface.
While it does not lead to the breakup of a hot strangelet but rather
favors the  creation of a cold strangelet, pion 
emission is suppressed at low temperature by a factor
$\sim exp(-m_{\pi}/T)$ ($\sim 0.01$ at $T$=30 MeV). Furthermore, for a 
strangelet with $E/A <$930 MeV, nucleon emission  proceeds via energy 
fluctuations and will be greatly suppressed \cite{gre87,rei88}.
If $E/A>$930 MeV, on the other hand, nucleon emission process 
has no threshold and goes much  faster, which might drive the hot strangelet to
a complete hadronization. 
Therefore, 
in this context we shall call a strangelet stable if its $E/A$ is lower
than 930 MeV throughout this work.

In Fig.~2(a) we have presented the (internal) energy per baryon as a function 
of temperature, for two different baryon numbers, $A=20$ and 100, with bag
constant $B^{1/4}$=145 MeV and strange quark mass $m_s$=150 MeV. Note that 
the strangelet with a baryon number $A$=100 is 
stable only 
for temperature $T \alt$20 MeV. 
(The dot-dashed line in Fig.~2(a) marks an energy per baryon of 930 MeV). 
We see also that the upper limit of
temperature for stable configuration of strangelets depends on baryon number.
This is a consequence of the inclusion of finite size effects.
More discussions will be given in Fig.~4.

Fig.~2(b)-(d) show the temperature dependence of the strangeness fraction $f_s$,
charge-to-baryon-number ratio $Z/A$, and 
averaged radius per baryon $R/A^{1/3}$ of strangelets with $A$=20, and 100,
for $B^{1/4}$=145 MeV and $m_s$=150 MeV. It can be seen that $f_s$ and 
$R/A^{1/3}$ increase, and $Z/A$ decrease slowly, when the temperature rises up.
It is expected that $f_s$ will saturate to unity, and $Z/A$ will decrease to
zero, when the temperature is high enough, since at that stage the mass 
difference between up, down and strange quarks is unimportant, the system will 
tend to become flavor symmetric ($N_u=N_d=N_s$) and globally charged neutral.
Our observation of $Z/A$ as a decreasing function of temperature differs from
that of Chakrabarty \cite{cha93}. 

In Fig.~3(a)-(d) we have plotted the energy per baryon $E/A$, strangeness 
fraction $f_s$, charge-to-baryon-number ratio $Z/A$, and averaged radius per
baryon $R/A^{1/3}$ as a function of baryon number $A$, for two temperatures
$T$=0 and 30 MeV, with $B^{1/4}$=145 MeV and $m_s$=150 MeV as well. One can
see from Fig.~1(a) that there is a lower limit $A_{cri}$ to the baryon 
number of a stable strangelet (whose $E/A<$930 MeV). This critical baryon
number is a function of temperature, parameters $B$ (bag constant) and $m_s$
(strange quark mass), as will be shown later. Except for sufficiently low
baryon number, we see that $f_s$, $Z/A$~and $R/A^{1/3}$~depends weakly 
on $A$. Their
typical values are $f_s\simeq$0.4-0.6, $Z/A\simeq$0.15-0.3, $R/A^{1/3}\simeq$
0.95-1 fm.

Fig.~4 shows the critical baryon number as a function of temperature, 
for two different values of $m_s$, $m_s$=150 MeV (full line) and $m_s$=0 
(dashed line)
 with $B^{1/4}$=145 MeV. For temperatures low enough, the critical 
baryon number increases little. However, beyond some temperature (about 
30 MeV for $m_s$=150 MeV, and 42 MeV for $m_s$=0) 
even bulk strange quark matter ($A\rightarrow
\infty$) could not be stable.

Before any concrete statements can be made, it is reasonable to discuss the 
parameter dependence. Next we study how the properties of strangelets depend
on the two main parameters in liquid drop model, bag constant $B$ and strange
quark mass $m_s$.

The variation with $B^{1/4}$~are plotted in Fig.~5(a)-(d) for strangelets with
a baryon number $A$=50, at temperature $T$=0 (full lines) and 30 MeV (dashed 
lines), with $m_s$=150 MeV. One can see from Fig.~5(a) that the energy per baryon
$E/A$ is proportional to 
$B^{1/4}$~for a given baryon number and temperature. While
the strangeness fraction $f_s$~increases with $B^{1/4}$, 
the charge-to-baryon-number ratio $Z/A$~
and averaged radius per baryon
$R/A^{1/3}$~are decreasing functions of $B^{1/4}$.

In Fig.~6 the critical baryon number $A_{cri}$ is plotted as a function of 
$B^{1/4}$, with the strange quark mass $m_s$=150 MeV. It can be seen that for  
sufficiently large bag constant, there could be no stable strange quark matter.
This upper limit of 
$B^{1/4}$~depends on temperature, $B^{1/4}\alt$153 MeV at $T$=0,
$B^{1/4}\alt$ 151 MeV at $T$=15 MeV,
and $B^{1/4}\alt$145 MeV at $T$=30 MeV. For strange quark mass less than 
150 MeV, the upper limit of $B^{1/4}$~moves up.

Fig.~7(a)-(d) show the variation with $m_s$ of properties of strangelets at
temperature $T$=0 and 30 MeV, with baryon number $A$=50 and $B^{1/4}$=145 MeV. 
For strangelets at $T$=0, both
the energy per baryon $E/A$ and charge-to-baryon number ratio 
$Z/A$ increase with increasing $m_s$, and
saturate when $m_s$ reaches the chemical potential $m_s\simeq \mu_s$. The 
strangeness fraction $f_s$ is a decreasing function of $m_s$, and goes to zero
when $m_s\alt \mu_s$. For finite temperature, $T$=30 MeV, the saturation
happens at a larger value of $m_s$. However, the averaged radius per baryon
$R/A^{1/3}$~depends little on $m_s$.

We have plotted in Fig.~8 the critical baryon number $A_{cri}$ as a function
of $m_s$, with 
$B^{1/4}$=145 MeV. Note that at zero temperature $A_{cri}$ increases
with increasing $m_s$, and suddenly becomes a constant at $m_s \simeq \mu_s$,
indicating the fact that for $m_s \alt \mu_s$, no strange quarks exist (see
Fig.~7(b) where $f_s\rightarrow 0$), and an increased value of $m_s$ does not
lead to an increase of strangelet energy. 

\section{Discussions and conclusions}
In the present work we have neglected QCD radiation corrections, as argued 
in Refs.\cite{far84,ber87} that an increased value of the strong coupling
constant $\alpha_s$ could be absorbed into a decrease in the bag constant
$B$. Coulomb effects could be included in liquid drop model self-consistently
as shown by Madsen \cite{mad93}. We show also in Fig.~1 
by the dashed lines the free energy per
baryon versus strangeness fraction after the inclusion of Coulomb energies.
One sees that Coulomb effects are rather negligible for the strangelet energy,
but could cause a change of the quark composition of the strangelet (Note in 
Fig.~1 that the value of $f_s$ minimizing $F/A$ is shifted a little by the
inclusion of Coulomb effects). The resultant corrections to $Z/A$ are within
the uncertainties caused by parameters $B$ and $m_s$, however.

We have found in this work that the stability of strangelets at zero and
finite temperature depends strongly on parameters $B$, the bag constant, and
$m_s$, the strange quark mass. An increased value of $B$ or $m_s$ (for 
$m_s<\mu_s$) tends to destabilize the strangelet. For $m_s$=150 MeV, the
strangelet can be stable only for 
$B^{1/4} \alt$153 MeV at $T$=0, $B^{1/4} \alt$151 MeV at $T$=15 MeV,
 and $B^{1/4} \alt$145 MeV at $T$=30 MeV.
For $B^{1/4}$=145 MeV, the strangelet is stable only for 
$T\alt$30 MeV with $m_s$=150 MeV, and $T\alt$42 MeV with $m_s$=0. 
This can impose important restrictions on the possible existence of hot 
strange quark matter, {\it e.g.} strange stars in astrophysics. From the fact
that ordinary nuclei are composed of nucleons instead of up-down quark
matter we can obtain a lower
limit ($B^{1/4}\agt$145 MeV) of the bag constant \cite{mad93}.
For the most extreme choice of parameters, {\it i.e.} 
$B^{1/4}$=145 MeV and $m_s$=0
(see dashed line in Fig.~4), we find that strange quark matter at a 
temperature $T\agt$42 MeV cannot 
exist stable.

The critical baryon number $A_{cri}$, below which the 
strangelet cannot be stable, is an increasing function of $T$, $B$ and $m_s$.
For $m_s$=150 MeV and $B^{1/4}$=145 MeV, the critical baryon number is about
$A_{cri}\simeq$25 at $T$=0, and $A_{cri}\simeq$60 at $T$=20 MeV.

The strangeness fraction of the strangelet, $f_s$, is found to increase slowly
with the increase of temperature $T$ or baryon number $A$ 
(except for $A \alt$100).
However, $f_s$ as an increasing function of 
$B^{1/4}$~and a decreasing function of 
$m_s$, is sensitive to the values of 
$B^{1/4}$~and $m_s$. For $m_s$=150 MeV and
$B^{1/4}$=145 MeV, the typical 
values of $f_s$ are about 0.4-0.7. In terms of the
simple relationship $Z/A=(1-f_s)/2$, the charge-to-baryon-number ratio $Z/A$
has typical values about 0.15-0.3, for $m_s$=150 MeV and $B^{1/4}$=145 MeV.

Finally the averaged radius per baryon, 
$R/A^{1/3}$, is a slowly increasing function of
$T$ or $m_s$, and a slowly decreasing function of $A$, whereas it decreases 
fast with the increase of $B^{1/4}$. For $m_s$=150 MeV, 
$B^{1/4}$=145 MeV, 
$R/A^{1/3}$~ranges  
from 0.9 to 1.1 fm.

To conclude, we have shown in the present work that liquid drop model including
finite size effects provides a successful description of properties of 
strangelets at zero and finite temperature, and allows studies for a wide
range of parameters. It is possible to carry out 
further explorations of phase structure,
phase evolution, and possible survival of strangelets in ultrarelativistic
heavy ion collisions within the framework of liquid drop model.

\acknowledgments
The authors would like to thank Prof. R. K. Su and Dr. S. Gao 
for fruitful discussions. Y.B.H. is grateful to Prof. Jes Madsen for helpful
correspondence.

\appendix
\section*{}
Eqs.~(\ref{eq1}) and (\ref{eq2}) are given by minimizing the free energy
\begin{eqnarray}
F=\sum_{i=q,s}(\Omega_i+N_i \mu_i)+BV,
\end{eqnarray}
with respect to $V$, $N_q$ and $N_s$, under the constraint
\begin{eqnarray}
\sum_i N_i=3A
\end{eqnarray}
for fixed baryon number $A$.
For the sake of simplicity we omit the surface and curvature terms in $F$ in 
the following derivations.

Constructing an auxiliary function 
\begin{eqnarray}
F^{\prime}=\sum_{i=q,s}\left\{\Omega_i(V,\mu_i(V,N_i))+N_i\mu_i(V,N_i)\right\}
+BV+\lambda \left(\sum_{i=q,s}N_i-3A\right),
\end{eqnarray}
we can derive the constrained minimum from
\begin{eqnarray}
\label{a1}
\left({\partial F^{\prime} \over \partial V}\right)_{N_i}=0,
\end{eqnarray}
and 
\begin{eqnarray}
\label{a2}
\left({\partial F^{\prime} \over \partial N_i}\right)_V=0.
\end{eqnarray}

Since $(\partial \Omega_i / \partial \mu_i)_V=-N_i$, Eq.~(\ref{a1})
leads to 
\begin{eqnarray}
\left({\partial F^{\prime} \over \partial V}\right)_{N_i}
=\sum_i\left\{\left({\partial \Omega_i
\over \partial V}\right)_{\mu_i}
+\left({\partial \Omega_i \over \partial \mu_i}\right)_V
\left({\partial \mu_i \over \partial V}\right)_{N_i}
+N_i\left({\partial \mu_i \over \partial V}\right)_{N_i}\right\}+B=0,
\end{eqnarray}
or 
\begin{eqnarray}
\sum_i\left({\partial \Omega_i \over \partial V}\right)_{\mu_i}+B=0,
\end{eqnarray}
which is Eq.~(\ref{eq1}).

From Eq.~(\ref{a2}) we obtain
\begin{eqnarray}
\left({\partial F^{\prime} \over \partial N_i}\right)_V=
\left({\partial \Omega_i \over \partial \mu_i}\right)_V
\left({\partial \mu_i \over \partial N_i}\right)_V+\mu_i
+N_i\left({\partial \mu_i \over \partial N_i}\right)_V+\lambda=0,
\end{eqnarray}
or
\begin{eqnarray}
\mu_i+\lambda=0,
\end{eqnarray}
which implies 
\begin{eqnarray}
\mu_q=\mu_s.
\end{eqnarray}
This is Eq.~(\ref{eq2}).



\newpage
\begin{center}
{\bf FIGURE CAPTIONS}
\end{center}

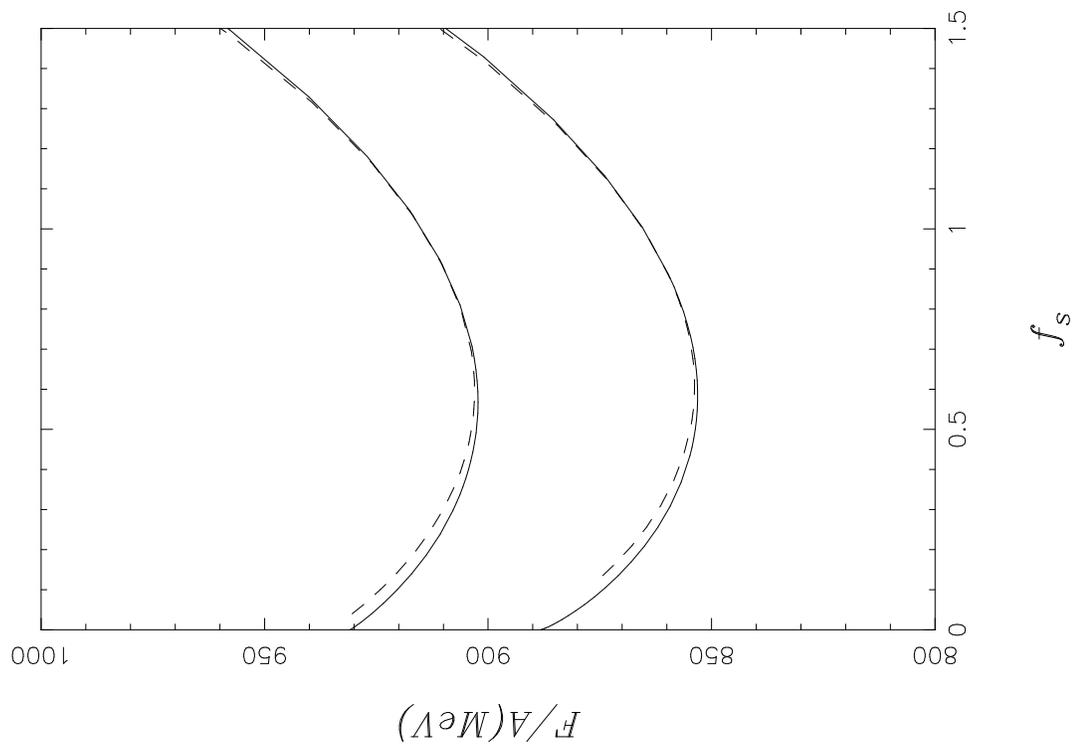
Fig.~1. Free energy per baryon $F/A$ as a function of the strangeness fraction 
$f_s$ with (dashed lines) and without (full lines) Coulomb corrections, at
temperature $T$=0 (upper two curves) and $T$=30 MeV (lower two curves), for
strangelets with a baryon number $A$=100, with $m_s$=150 MeV and $B^{1/4}$=145
MeV.

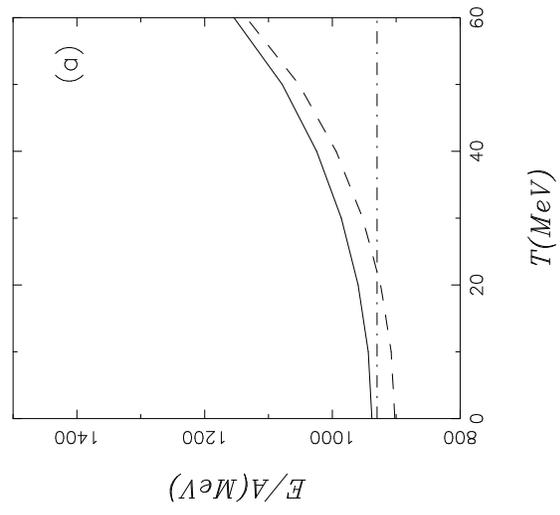
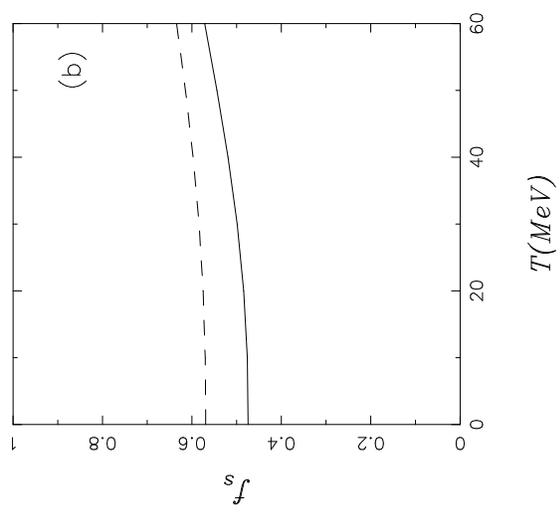
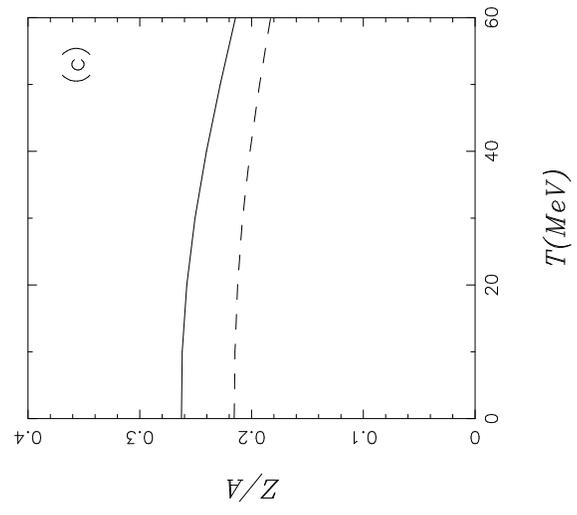
Fig.~2.(a)Internal energy per baryon $E/A$, (b)strangeness fraction $f_s$,
(c)charge-to-baryon-number ratio $Z/A$, and (d)averaged radius per baryon
$R/A^{1/3}$ as a function of temperature $T$, for $A$=20 (full lines)
and $A$=100 (dashed lines), with $m_s$=150 MeV and $B^{1/4}$=145 MeV.

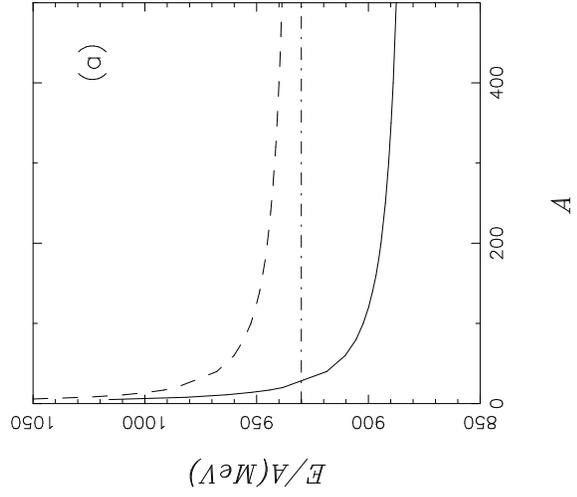
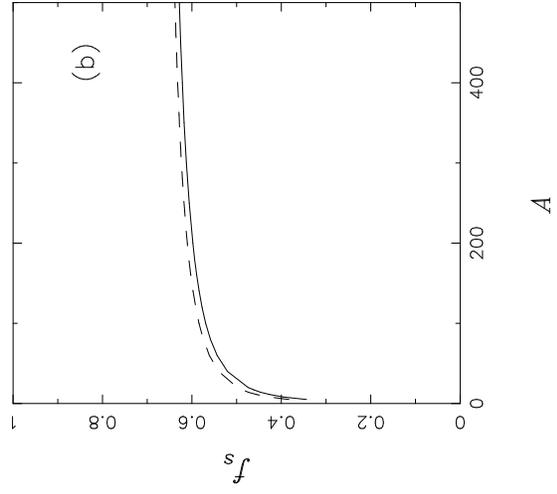
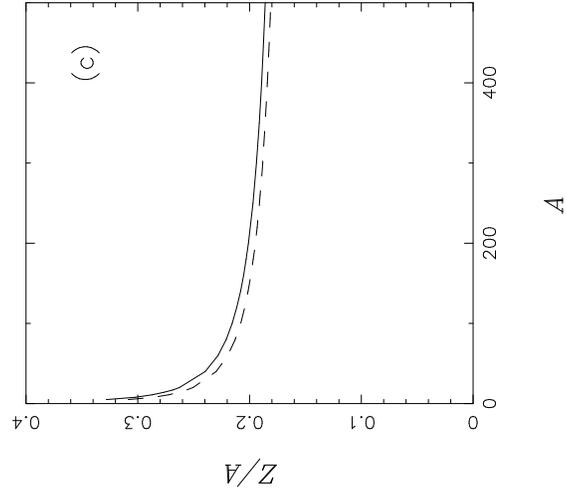
Fig.~3.(a)Internal energy per baryon $E/A$, (b)strangeness fraction $f_s$,
(c)charge-to-baryon-number ratio $Z/A$, and (d)averaged radius per baryon
$R/A^{1/3}$ as a function of baryon number $A$, at temperature $T$=0 (full
lines) and $T$=30 MeV (dashed lines), with $m_s$=150 MeV and $B^{1/4}$=145 MeV.

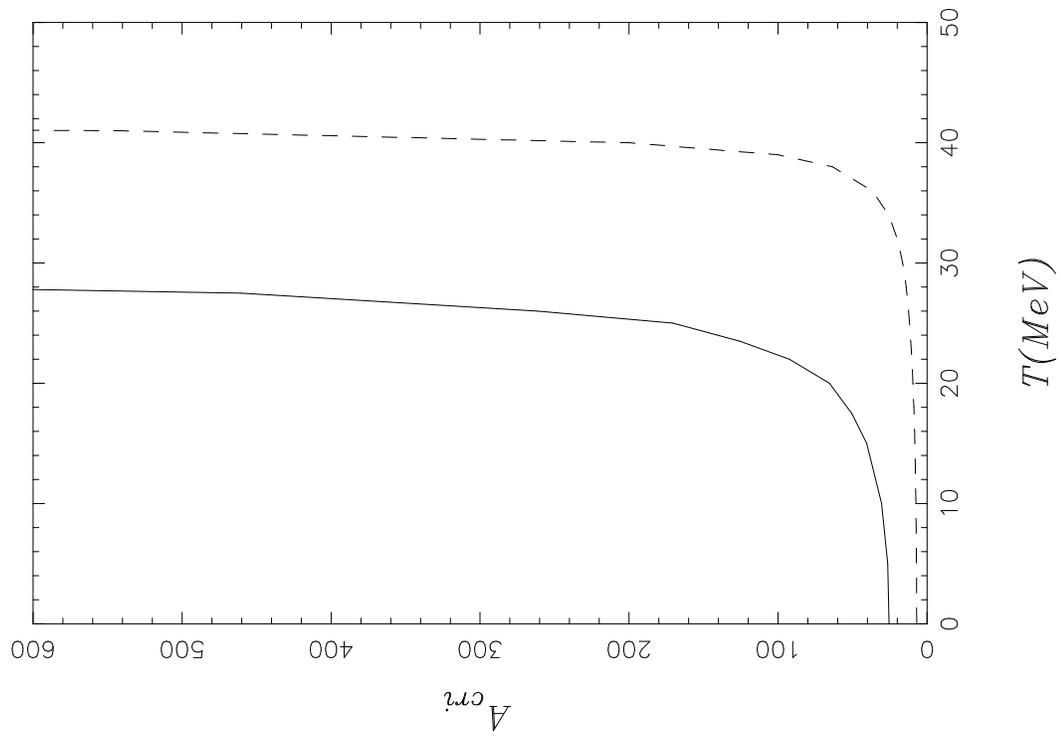
Fig.~4. Critical baryon number $A_{cri}$ as a function of temperature $T$,
for two different values of $m_s$, $m_s$=150 MeV (full line) and $m_s$=0 (dashed
line), with $B^{1/4}$=145 MeV.

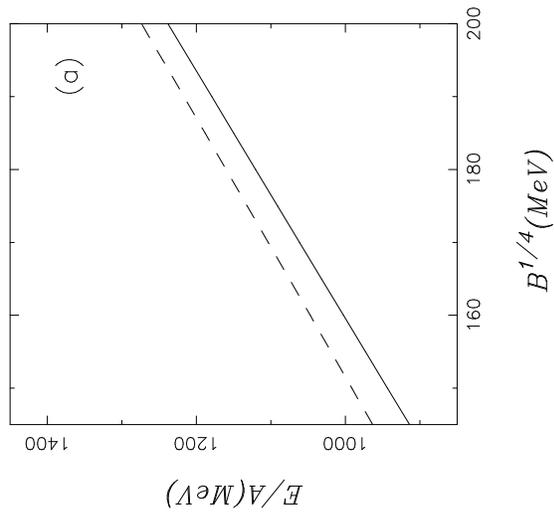
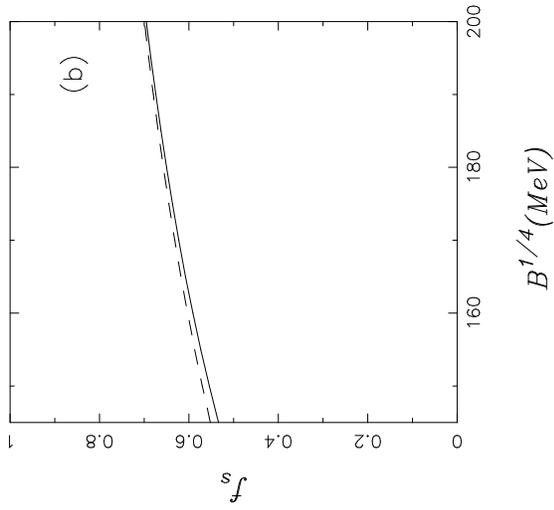
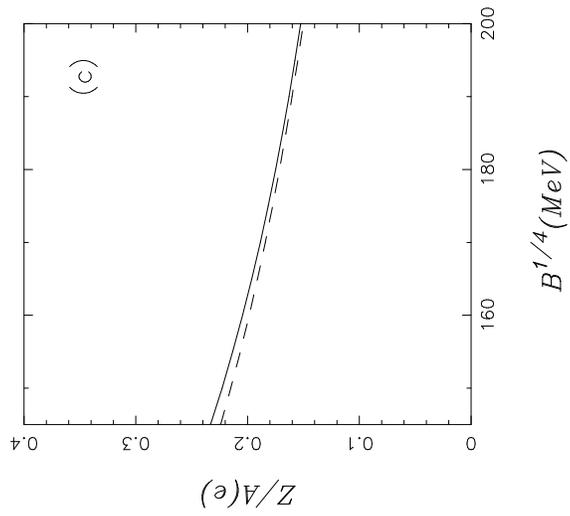
Fig.~5.(a)Internal energy per baryon $E/A$, (b)strangeness fraction $f_s$,
(c)charge-to-baryon-number ratio $Z/A$, and (d)averaged radius per baryon
$R/A^{1/3}$ as a function of bag constant $B^{1/4}$, for strangelets with baryon
number $A$=50, at temperature $T$=0 (full lines) and $T$=30 MeV (dashed 
lines), with $m_s$=150 MeV.

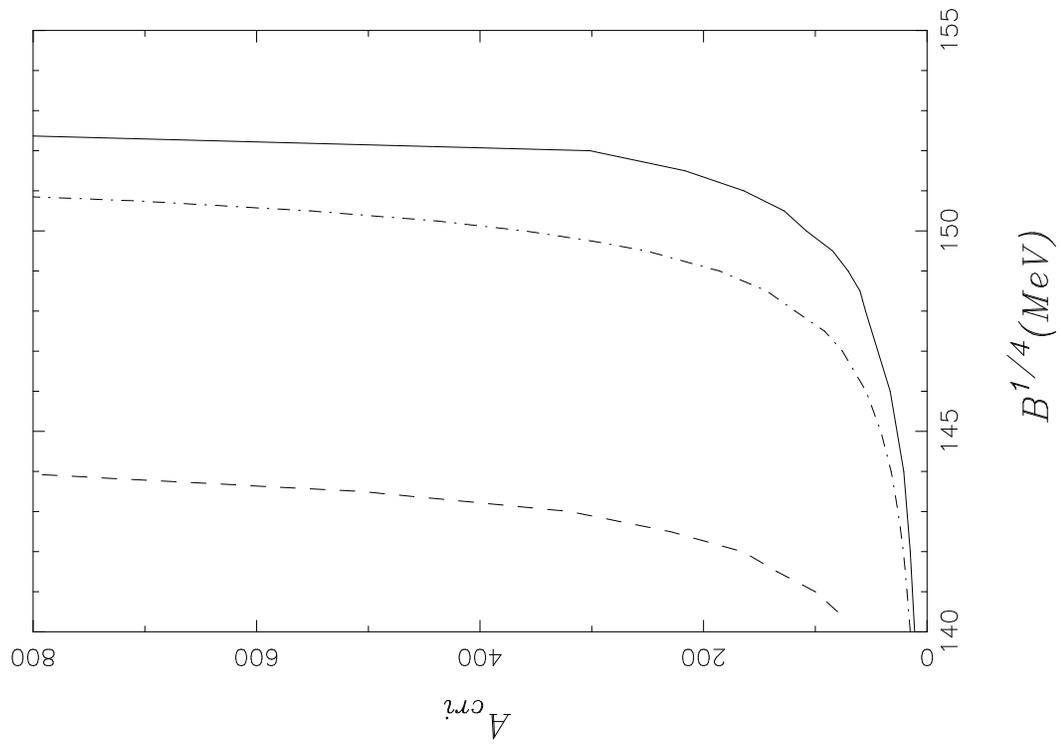
Fig.~6. Critical baryon number $A_{cri}$ as a function of bag constant 
$B^{1/4}$,
for strangelets at temperature $T$=0 (full line),
$T$=15 MeV (dot-dashed line), and $T$=30 MeV (dashed line), with $m_s$=150 MeV.

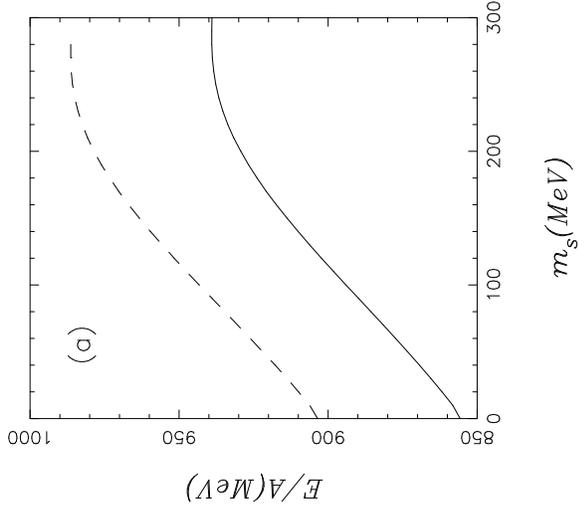
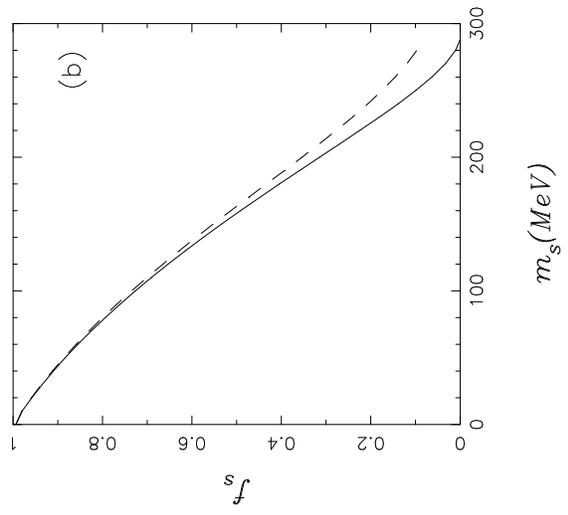
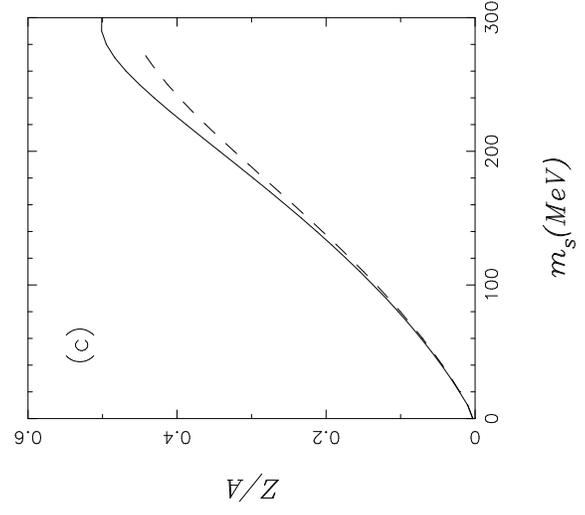
Fig.~7.(a)Internal energy per baryon $E/A$, (b)strangeness fraction $f_s$,
(c)charge-to-baryon-number ratio $Z/A$, and (d)averaged radius per baryon
$R/A^{1/3}$ as a function of strange quark mass $m_s$, for strangelets with 
a baryon number $A$=50, at temperature $T$=0 (full lines) and $T$=30 MeV
(dashed lines), with $B^{1/4}$=145 MeV.

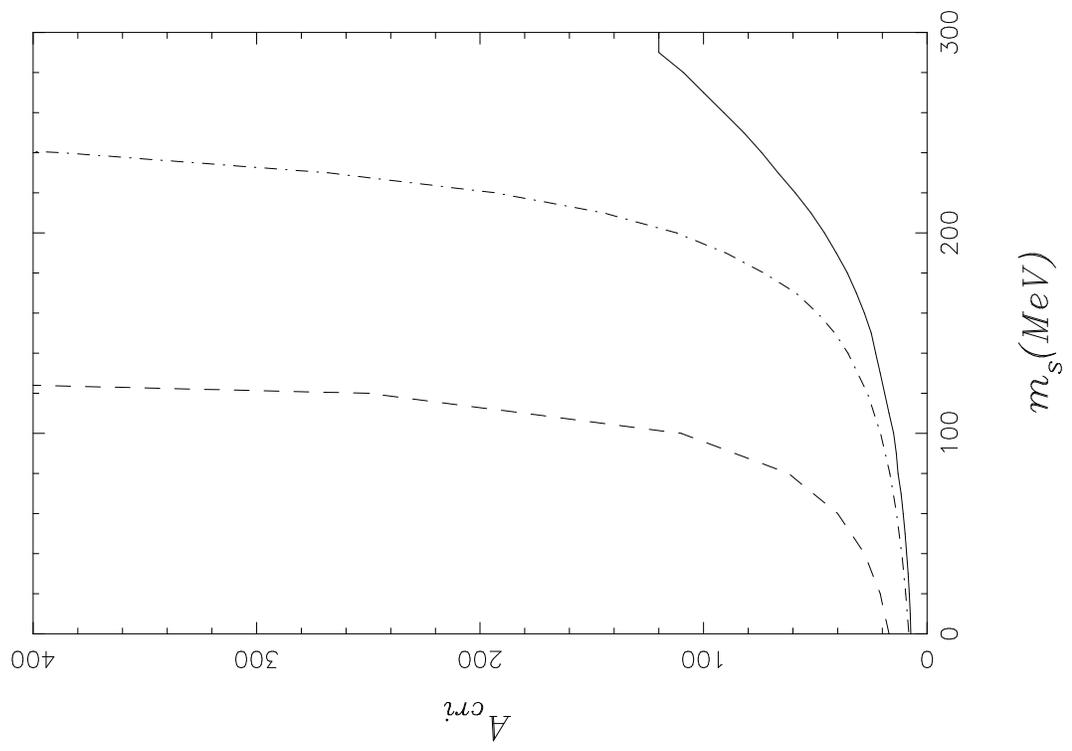
Fig.~8. Critical baryon number $A_{cri}$ as a function of strange quark mass 
$m_s$, for strangelets at temperature $T$=0 (full line) 
$T$=15 MeV (dot-dashed line) , and $T$=30 MeV
(dashed line), with $B^{1/4}$=145 MeV.
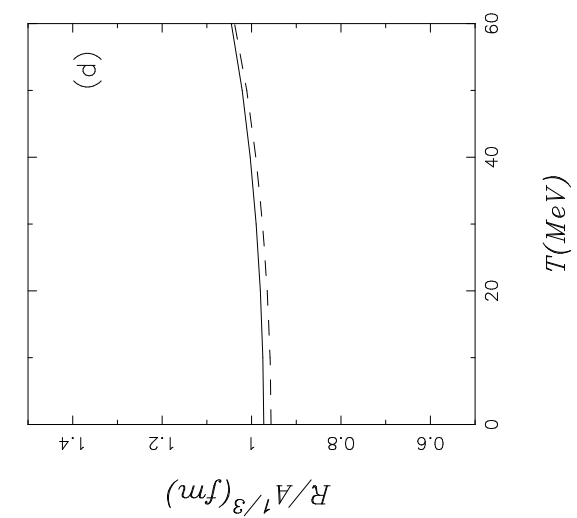
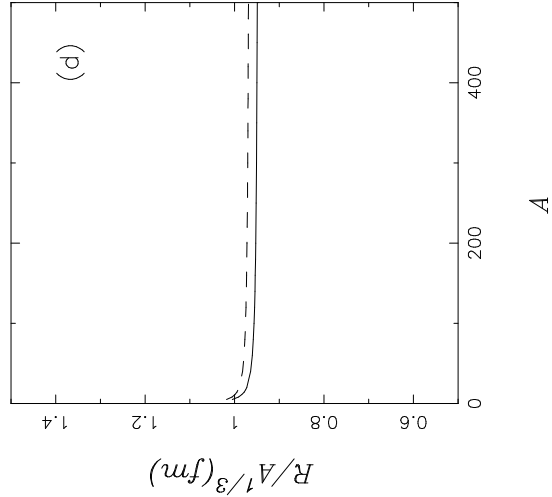
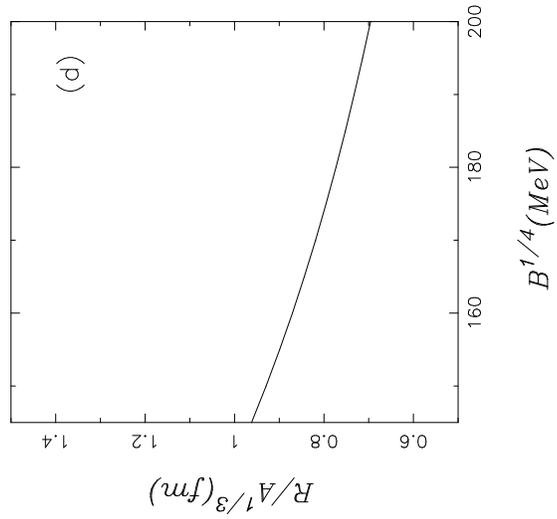
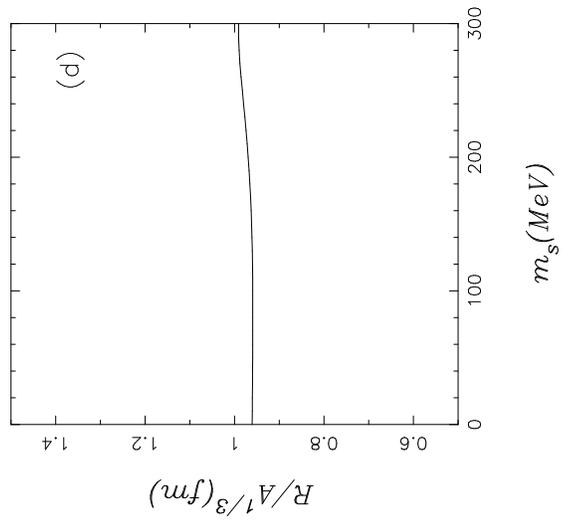
\end{document}